\documentclass[aps,prl,twocolumn,superscriptaddress]{revtex4-1}
\usepackage{graphicx}
\usepackage{amssymb}
\usepackage{epsfig}
\usepackage{graphicx}
\usepackage{amsmath}
\usepackage{array,color}
\usepackage{leftidx}
\usepackage{mathrsfs}
\usepackage{ulem}
\usepackage{hyperref}
\usepackage[dvipsnames]{xcolor}
\hypersetup
{colorlinks=true,
citecolor=blue,,
urlcolor=blue,
linkcolor=blue,
anchorcolor=blue,
filecolor=blue,}

\begin{document}
\title{Collisions of Majorana Zero Modes}
\author{Liang-Liang Wang}
%\email{wangliangliang@westlake.edu.cn}
\affiliation{School of Science, Westlake University, 18 Shilongshan Road, Hangzhou 310024, Zhejiang Province, China}
\affiliation{Institute of Natural Sciences, Westlake Institute for Advanced Study, 18 Shilongshan Road, Hangzhou 310024, Zhejiang Province, China}
\author{Wenjun Shao}
%\email{wangliangliang@westlake.edu.cn}
\affiliation{School of Science, Westlake University, 18 Shilongshan Road, Hangzhou 310024, Zhejiang Province, China}
\affiliation{Institute of Natural Sciences, Westlake Institute for Advanced Study, 18 Shilongshan Road, Hangzhou 310024, Zhejiang Province, China}
\author{Jian Li}
\email{lijian@westlake.edu.cn}
\affiliation{School of Science, Westlake University, 18 Shilongshan Road, Hangzhou 310024, Zhejiang Province, China}
\affiliation{Institute of Natural Sciences, Westlake Institute for Advanced Study, 18 Shilongshan Road, Hangzhou 310024, Zhejiang Province, China}

\date{{\small\today}}

\begin{abstract}
%====[Abstract]====
We investigate the collisions of Majorana zero modes, which are presented as inter-soliton collisional events in fermionic superfluids with spin-orbit coupling.
%====[Abstract]====
%Our results demonstrate that Majorana zero modes are not only topologically protected, but also self-protected.
%====[Abstract]====
Our results demonstrate that, the zero energy splitting, induced by the overlapping of inter-soliton Majorana wave-functions upon collision, generates an effective repulsive force for Majorana states, which in turn protected themselves against into bulk excitation.
%====[Abstract]====
As a result, the collision between solitons associated with Majorana zero modes appears to be repulsive and elastic, as they do not penetrate each other but instead repel without energy loss.
%====[Abstract]====
%This is distinct from other repulsion mechanisms considered previously.
%====[Abstract]====
As well, similar repulsive behavior is observed in collisions between soliton-induced and defect-pinned Majorana zero modes.
%====[Abstract]====
Our research offers new insights into the features of Majorana fermions, and robustness in the collisions of Majorana zero modes bodes well for the prospects of topological quantum computation with a multitude of Majorana qubits.

\end{abstract}
%\pacs{67.85.-d, 03.75.Ss, 05.30.Fk}
\maketitle

\textit{Introduction.}--
%====[Intro-01]====
Majorana zero modes (MZMs)\cite{Majorana-1937,WilczekF-2009,BrouwerPW-2012,ElliottSR-2015}are exotic, neutral quasiparticles composed of the equivalent contributions of the particle and the hole.
%====[Intro-01]====
They are of fundamental scientific importance and could have profound technological applications for fault-tolerant quantum computation\cite{HasanM-2010,QiX-2011,TewariS-2007,NayakC-2008}, quantum memory \cite{KitaevAY-2001,KitaevAY-2003,ShorPW-1995,BiercukMJ-2009,MaurerPC-2012}and quantum random-number generation\cite{RarityJG-1994,DongLing-2013}.
%====[Intro-01]====
Similar to the Bardeen-Cooper-Schrieffer (BCS) pairing mechanism between electron creation and annihilation\cite{BCS-1957}, the mixing of the particle and hole results in the ground states doubly degenerate, and two MZMs then form a protected qubit that is not locally measurable\cite{KitaevAY-2001}.
%====[Intro-01]====
These non-Abelian quasiparticles are believed to merge in topological superconducting and superfluid systems, including the interfaces of \textit{s}-wave superconductor and topological insulator\cite{FuL-2008,LutchynRM-2010,OregY-2010,AliceaJ-2010,JiangL-2011,MourikV-2012}, intrinsic two-dimensional superconductors with \textit{p}-wave pairing symmetry\cite{ReadN-2000,MizushimaT-2008,BjornsonK-2015,MurrayJ-2015}, as well as topological ferromagnetic metal chains\cite{PergeS-2013,LiJ-2014,PawlakR-2016}.
%====[Intro-01]====
After the realization of spin-orbit coupling (SOC) in ultracold gases\cite{Lin-2011,Wang-2012,Wu-2016,Meng-2016,Huang-2016}, atomic fermionic superfluids enter the topological state, offering a disorder-free and highly controllable platform for studying Majorana physics\cite{Dalibard-2011,Wu-2013,Devreese-2014,Zhai-2015}.

%====[Intro-02]====
MZMs appear within the cores of certain soliton excitations\cite{Xu-2014,Liu-2015,Mateo-2022}, where a phase kink across the dip-like structure of the order parameter arises from the atomic phase imprinting techniques\cite{Denschlag-2013, Burger-1999, Anderson-2000, Ku-2016}.
%====[Intro-02]====
These zero-energy states lead to the degeneracy of the many-body ground states, which hinges on the precise degeneracy of the MZMs in various soliton cores.
%====[Intro-02]====
Physically, soliton is a good candidate for the control and manipulation of Majorana qubits owing to its classical particle-like character\cite{Kartashov-2011}, and networks of such topological excitations have been proposed for the progress of quantum computing\cite{El-2005, Tecas-2013, Shaukat-2017, Ezawa-2020}.
%====[Intro-02]====
Nevertheless, in the presence of multiple solitons, inter-soliton colliding events become possible\cite{ScottR-2012} and overlapping between Majorana states within different solitons are expected to lift the Majorana state degeneracy to some degree, even may lead to a complete breakdown.
%====[Intro-02]====
For the purposes of topological quantum computation, it is vital to figure out the stability of the MZMs in colliding processes.

\begin{figure*}[t]
\centering
\includegraphics[width=0.95\textwidth]{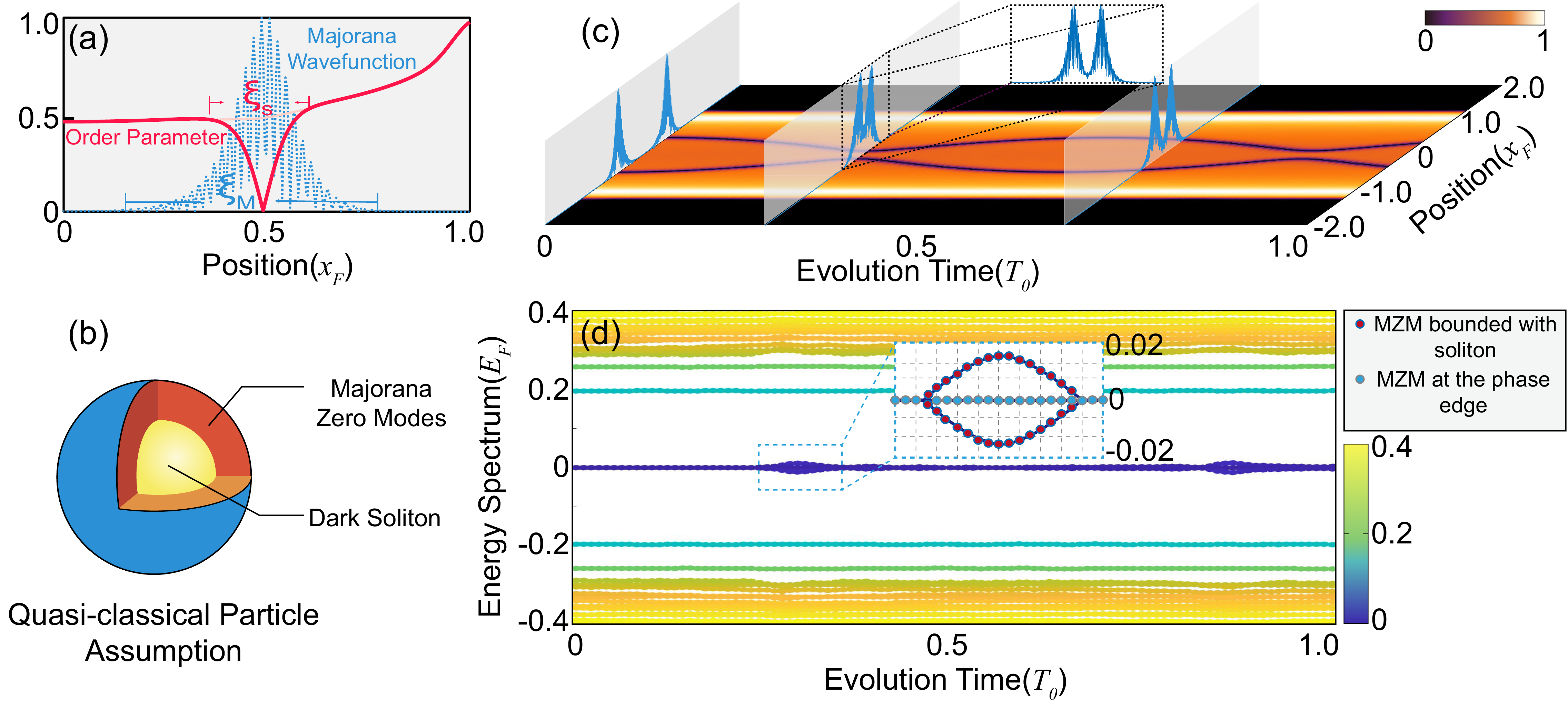}\\
\caption{(color online) 
%Collisions of Majorana zero modes. 
(a) The solitonic pairing order parameter $\Delta(x)/E_{F}$(solid red curve) and corresponding wave-function $|u_{\uparrow}(x)|$ (dashed blue curve) of the lowest-energy Majorana zero states inside the dark soliton at $x_{0}=0.5x_{F}$. (b) Quasi-classical illustration of the two-shell structure: dark soliton surrounded by Majorana wave-functions. (c) The evolution of the order parameter profile of the superfluid, while sequential snap shots of the lowest energy Majorana wave-functions (specifically $|u_{\uparrow}(x,t)|$) are displayed above it. The solitons begin at rest at a distance $\pm 0.5x_{F}$ from the trap centre. (d) Corresponding plot of the time dependent energy spectrum $E_{\eta}(t)$. Other parameters are $\alpha k_{F}=E_{F}$, $h_{z}=1.0E_{F}$.
}
\label{Fig:Soliton Collision}
\end{figure*}

%====[Intro-03]====
In this Letter we address this question by observing soliton collisions in one-dimensional fermionic superfluids with SOC.
%====[Intro-03]====
The existence of MZMs within the solitons has important consequences for the physics of soliton collisions.
%====[Intro-03]====
Based on our numerical simulations with time-dependent Bogoliubov-de Gennes equation, we find that the soliton collision with MZMs appears repulsive and completely elastic, i.e., they do not penetrate each other but instead repel with a well distance that the soliton matters can be safely regarded as untouched.
%====[Intro-03]====
Thus the soliton matter interaction could be negligible and allow us to clearly identify the collisional nature of MZMs.
%====[Intro-03]====
We confirm that the overlapping of inter-soliton Majorana wave-functions upon collision lift the Majorana degeneracy from zero.
%====[Intro-03]====
Moreover, the energy splitting of the Majorana states creates an effective repulsive force for MZMs, which in turn protected themselves against into bulk excitation.
%====[Intro-03]====
We stress that this is distinct from other repulsion mechanisms considered previously.
%====[Abstract]====
And, for a contrast, we also demonstrate that soliton collisions become increasingly inelastic as we tune the system into the topological trivial phase.
%====[Intro-03]====
We further reexamine the repulsive interaction in collisions between soliton-bound and defect-pinned Majorana states.
%====[Intro-03]====
Our results show that Majorana zero modes are not only topologically protected, but also self-protected.
%====[Intro-03]====
This provides new insights into the intrinsic nature of Majorana states and these unusual colliding properties may open an alternative way to detect and discriminate Majorana qubits for fault-tolerant topological quantum computations.

\textit{Theoretical model.} --
%====[Model-01]====
Our starting point is the description of dark soliton-collisions in one-dimensional spin-half fermionic superfluids with SOC.
%====[Model-01]====
The one-dimensional setting ensures the dark solitons stable with respect to the snake instability\cite{Brand-2002,Ku-2016} and the SOC effect can be realized by two counter-propogating Raman lasers\cite{Wang-2012}.
%====[Model-01]====
Within the standard mean-field framework, the dynamics of a fermionic superfluid can be modeled by the well-known time-dependent Bogoliubov-de Gennes (TDBdG) equation, whose one-dimensional form reads
\begin{equation}
\label{main BdG equation}
\begin{bmatrix}
H_{0} & \Delta(x, t) \\
\Delta^{\ast}(x, t) & -\sigma_{y}H_{0}^{\ast}\sigma_{y}\\
\end{bmatrix}
\Phi_{\eta}(x, t)=
i\hbar\frac{\partial}{\partial t}
\Phi_{\eta}(x, t)
,
\end{equation}
where the wave-functions is $\Phi_{\eta}\equiv [u_{\uparrow, \eta}, u_{\downarrow, \eta}, v_{\downarrow, \eta}, -v_{\uparrow, \eta}]^{T}$ in the Nambu representation.
%====[Model-01]====
The single particle grand-canonical Hamiltonian has the form $H_{0}=-\hbar^2\partial^{2}_{x}/2m+m\omega^2x^2/2-i\alpha\hbar \partial x \sigma_{y}-\mu+h_{z}\sigma_{z}$, describing the motion of fermionic atoms confined in a harmonic trapping potential with an oscillation frequency $\omega$.
%====[Model-01]====
$\alpha$ is the SOC strength, $h_{z}$ the effective Zeeman filed, and $\mu$ the chemical potential.
%====[Model-01]====
The order parameter has the self-consistent form $\Delta(x, t) =g_{\mathrm{1D}}\sum_{\eta}[u_{\uparrow, \eta}(x,t)v_{\downarrow, \eta}^{\ast}(x,t)f(-E_{\eta})-u_{\downarrow, \eta}(x,t)v_{\uparrow, \eta}^{\ast}(x,t)f(E_{\eta})]$
and the atom density function is given by $n_{\sigma}(x,t)=\sum_{\eta}|v_{\sigma, \eta}(x,t)|^{2}f(-E_{\eta})+ |u_{\sigma, \eta}(x,t)|^{2}f(E_{\eta})$, where $g_{1D}=-2\hbar^2/(ma_{1D})$ is the effective interatomic coupling given by an $s$-wave scattering length $a_{1D}$.
%====[Model-01]====
$f(E)=1/[e^{E/k_{B}T}+1]$ is the Fermi-Dirac distribution at a temperature $T$, and the summation is over the quasi-particle state $E_{n}\geqslant 0$.
%====[Model-01]====
Eq.(\ref{main BdG equation}) should be calculated self-consistently with the constraints of a fixed total atomic number $N=\int dx[n_{\uparrow}(x,t) + n_{\downarrow}(x,t)]$ and the above definition of the order parameter.

%====[Model-02]====
The particle-hole symmetry of BdG Hamiltonian ensures that MZMs only occur at exact zero energy and typically localized in the vicinity of defects, such as phase edges or solitons.
%====[Model-02]====
%====[Model-02]====
In this work, multiple dark solitons may be chosen to be \textit{real} and $\pi$-phase kinks are introduced at the spots $\{x_{i}, i=1,2,3,...\}$, given by
\begin{equation}
\Delta(x)=|\Delta(x)|\exp[i\pi \sum_{i}\Theta(x-x_{i})],
\label{Initial guess}
\end{equation}
where $\Theta(x)$ is the Heaviside step function.
%====[Model-02]====
We seek the stationary solutions self-consistently with the use of Eq. (\ref{Initial guess}), after a number of iterations up to convergence.
%====[Model-02]====
When the local superfluid becomes topological\cite{Xu-2014}, a pair of MZMs appear inside the soliton core and it is useful to note that the interaction between MZMs inside a dark soliton vanishes due to the intrinsic property of the dark soliton: a sharp phase jump.
%====[Model-02]====
In dynamic simulations, we use a dimensionless interaction parameter to characterize the interaction strength, $\gamma=-mg_{1D}/\hbar^2 n$, which is basically the ration between the interaction and kinetic energy at the density $n$.
%====[Model-02]====
We choose the Fermi vector and energy, $k_{F}=\pi n/2$ and $E_{F}=\hbar^2k_{F}^2/2m$, as the units of wave-vector and energy, respectively.
%====[Model-02]====
In a trapped cloud with $N$ atoms, it is convenient to use the peak density of a non-interacting Fermi gas in the Thomas-Fermi approximation at the trap center, $n=(2/\pi)\sqrt{Nm\omega/\hbar}$, although the Fermi cloud itself is an interacting gas.
%====[Model-02]====
Throughout the work, we consider only zero temperature and take the interaction parameter $\gamma=\pi$, the total atomic number $N=100$.

\begin{figure}[t]
\centering
\includegraphics[width=0.45\textwidth]{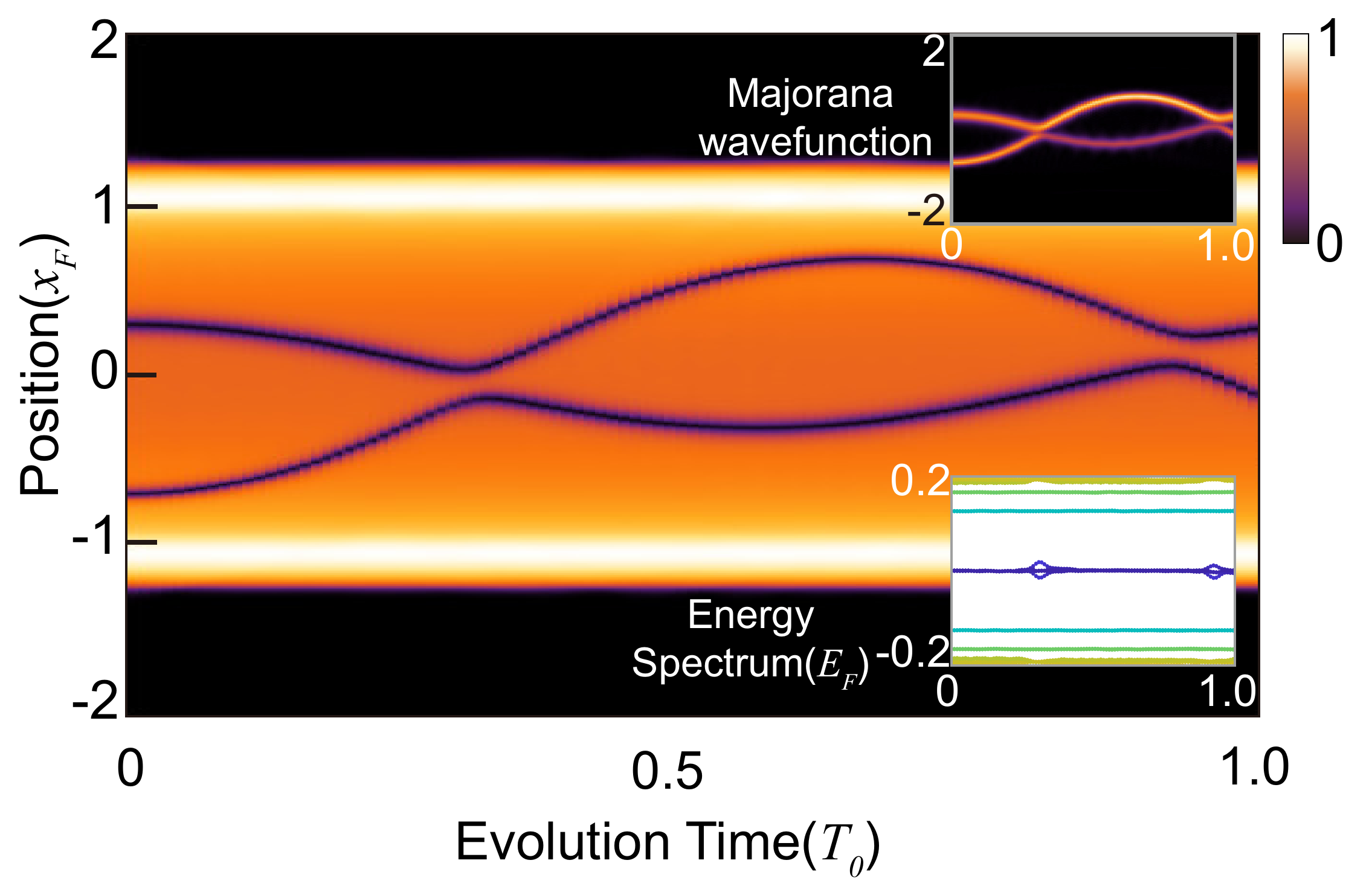}\\
\caption{(color online) Example of the evolution of an asymmetrical collision with two solitons prepared at $-0.8x_{F}$ and $0.3x_{F}$, respectively. The top inset shows the spatiotemporal contour plot of the Majorana wave-function, and bottom inset gives corresponding time dependent energy spectrum, where we can also observe the energy splitting in the collision.}
\label{Fig: Asymmetrical Collision}
\end{figure}

\textit{Collisions of Majorana zero modes.}--
%====[Soliton Collision-01]====
We first prepare the system in the topological phase and then place two dark solitons with a symmetrical displace, to observe the \textit{head-on} collision at the trap centre.
%====[Soliton Collision-01]====
Majorana states emerge around the nodal point of the soliton, which is in contrast to bosonic superfluids, where the Gross-Pitaevskii solitons are structureless.
%====[Soliton Collision-01]====
The initial soliton separation $x_{i}=\pm 0.5x_{F}$ is large enough that the Majorana wave-function overlapping is negligible.
%====[Soliton Collision-01]====
%walk-off diverge of the trajectories indicates the presence of repulsive forces.
% healing lengths, which defines the range of the repulsive soliton interaction. 
Before the collision, a close inspection of the stationary dark soliton at $0.5x_{F}$ is shown in Fig.(\ref{Fig:Soliton Collision}a).
%====[Soliton Collision-01]====
Owing to the appearance of MZMs within the soliton, the inter-soliton effects should be twofold: first, the mere soliton matter interaction arises sharply at spacings of the order of the soliton core width $\xi_{S}$, and second, the soliton-induced Majorana wave-function overlap with the length scale $\xi_{M}$, which describes the MZM behavior like $\sim\cos{(\pi k_{F}x/2)\exp(-x^2/\xi_{M}^2)}$.
%====[Soliton Collision-01]====
$\xi_{M}$ is larger than $\xi_{S}$ ($\xi_{M}\approx 2.5\xi_{S}$ in this case) and then we predict the Majorana soliton possesses a two-shell structure: soliton matter core surrounded by Majorana wave-functions (shown in Fig.(\ref{Fig:Soliton Collision}b)).
%====[Soliton Collision-01]====
The collisions should be fascinating due to their multi-component nature.

%====[Soliton Collision-01]====
In Fig.(\ref{Fig:Soliton Collision}c), we present the evolution of the order parameter profile of the superfluid, while sequential snapshots of the lowest energy Majorana wave-functions (specifically $|u_{\uparrow}(x,t)|^2$) are displayed above it.
%====[Soliton Collision-01]====
Clearly, the resultant collision appears to be repulsive and elastic: the solitons slow down as they approach one another, come to a halt with a well distance of $\xi_{\mathrm{min}}\approx0.3x_{F}$, and then reflect back to their original positions.
%====[Soliton Collision-01]====
With ($\xi_{M}>\xi_{\mathrm{min}}>\xi_{S}$), the soliton cores remain untouched and the Majorana wave-functions overlap (shown in Fig.(\ref{Fig:Soliton Collision}c)).
%====[Soliton Collision-01]====
The soliton matter interaction could be negligible and thus we safely attribute the repulsive force just to the Majorana states, allowing us to clearly identify the nature of interactions between MZMs.
%====[Soliton Collision-01]====
To grasp the main physics, the time dependent energy expectations $\left<E_{\eta}(t)\right>=\left<\Phi_{\eta}(x, t)|H_{\mathrm{BdG}}(x,t)|\Phi_{\eta}(x, t)\right>$ are calculated to observe the energy splitting of zero modes in the collision, as shown in Fig. (\ref{Fig:Soliton Collision}d).
%====[Soliton Collision-01]====
It's worth noting that, far away from each other, MZMs always have zero energy irrespective of the soliton velocity.
%====[Soliton Collision-01]====
As solitons get close enough, the zero modes split into a pair of levels $(E_{0}, -E_{0})$, which is proportional to the Majorana wave-function overlapping.
%====[Soliton Collision-01]====
This energy upshift creates a repulsive force for solitons that block their appropinquity and in turn drastically protects itself against scattering into bulk states.
%====[Soliton Collision-01]====
That's why the two solitons exhibit elastic collision only.

%====[Soliton Collision-01]====
%To cross check the above-made statements
Moreover, asymmetrical collision is shown in Fig.(\ref{Fig: Asymmetrical Collision}).
%====[Soliton Collision-01]====
During collision, solitons exchange energy via the Majorana wave-function overlapping, hence, solitons with different velocities (increase the soliton speed by increasing $x_{i}$), the quantity of motion is preserved, i.e., the low-speed soliton after collision propagate with high velocity, while the high-speed one runs slowly, looks like passing through one another without changing form.
%====[Soliton Collision-01]====
%所有的数值结构都表明：
The observed behavior of the above collision reveal that the Majorana states are not only topological protected, but also self-incurred protected, an aspect of the Majorana state that hadn't been explored before.
%====[Soliton Collision-01]====
This can ensure a more robust multi-Majorana quasiparticles transport.
% Therefore, the measurement reveal an unexplored aspect of the Majorana state
%Based on the above investigations,
%we claim that Majorana quasiparticles are not only topological protected, but also self-incurred protected.
%====[Soliton Collision-01]====
%====[Soliton Collision-01]====
%Here we note that the constant zero modes are the MZMs at the phase edges.
% It is worthwhile mentioning that the constant zero modes are the MZMs at the phase edges.

%show similar patterns (as elaborated on in Supplementary Material).
\begin{figure}[t]
\centering
\includegraphics[width=0.48\textwidth]{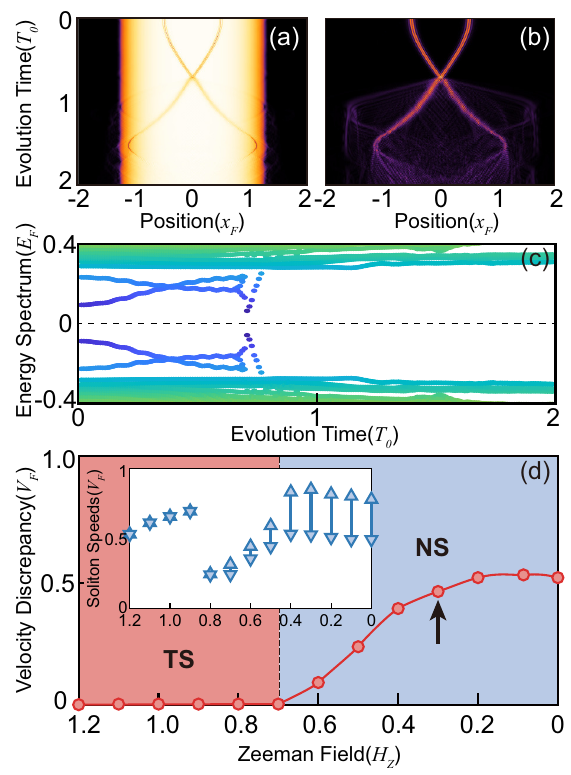}\\
\caption{(color online)
(a-b) gives the collision of solitons in normal superfluid (NS) phase with $h_{z}=0.3E_{F}$, where the left column (a) is
 the paring order parameter $\Delta(x,t)$, and the right one (b) shows the evolution of the lowest energy Andreev state (specifically $|u_{\uparrow}|$).
The collision is inelastic and the lost energy is converted into small amplitude ripples (shown both in (a) and (b)).
As a result, soliton-induced Andreev states is up-shifted into the bulk quasi-particle scattering continuum, which is shown in (c).   
(d) The velocity discrepancy $\Delta v$ between soliton velocity immediately before and after the first collisions as a function of Zeeman field $h_{z}$.
In the normal superfluid (NS) phase, the velocity discrepancy increases from zero, denoting the collision is getting inelastic.
As shown in the inset, blue downward (upward) pointing triangles are the soliton velocities before (after) the first collision.
}
\label{Fig: Normal Soliton Collision}
\end{figure}

%====[Soliton Collision-02]====
%To gain further intuition
Next, for a contrast, we tune the system into the topological trivial regime and observe the properties of the corresponding soliton collision.
%====[Soliton Collision-02]====
In Fig.(\ref{Fig: Normal Soliton Collision}), under the threshold $h_{z}\approx0.7E_{F}$, the superfluid enter the topological trivial phase and finite energy Andreev-like bound states emerge near the point node of the soliton.
%====[Soliton Collision-02]====
Observably, inelastic soliton collision is presented and the lost energy is converted into small-amplitude density ripples that emanate from the point of the collision (shown in Fig.(\ref{Fig: Normal Soliton Collision}a)). 
%====[Soliton Collision-02]====
Counter-intuitively, the dark solitons with lower energy have a higher velocity and vice versa, the soliton actually move faster after losing energy.
%====[Soliton Collision-02]====
Fig. (\ref{Fig: Normal Soliton Collision}b) shows the corresponding wave-function of lowest Andreev bound state (specifically $|u_{\uparrow}(x,t)|^2$).
%====[Soliton Collision-02]====
One can clearly see the corresponding ripples in the amplitude and, as a result, soliton-induced Andreev bound states are strongly up-shifted into the bulk quasi-particle scattering continuum, which eventually cause an emission of the sound and loss of the soliton energy (see Fig.(\ref{Fig: Normal Soliton Collision}c)).
%====[Soliton Collision-02]====
Furthermore, to quantify the elasticity of the collision, we evalued the soliton speed discrepancy immediately before and after the first collisions with the decreasing Zeeman field in Fig.(\ref{Fig: Normal Soliton Collision}d).
%====[Soliton Collision-02]====
The phase diagram for occurrence of elastic and inelastic collisions is obtained, which strongly depends on the topological properties.
%====[Soliton Collision-02]====
The speed discrepancy rises suddenly around the transition point and the collisions become increasingly inelastic in topological trivial phase.
%====[Soliton Collision-02]====
The different collision properties between Majorana and Andreev solitons may lead to an interesting technique of soliton filter for distinguishing the Majorana states.

\begin{figure}[t]
\centering
\includegraphics[width=0.45\textwidth]{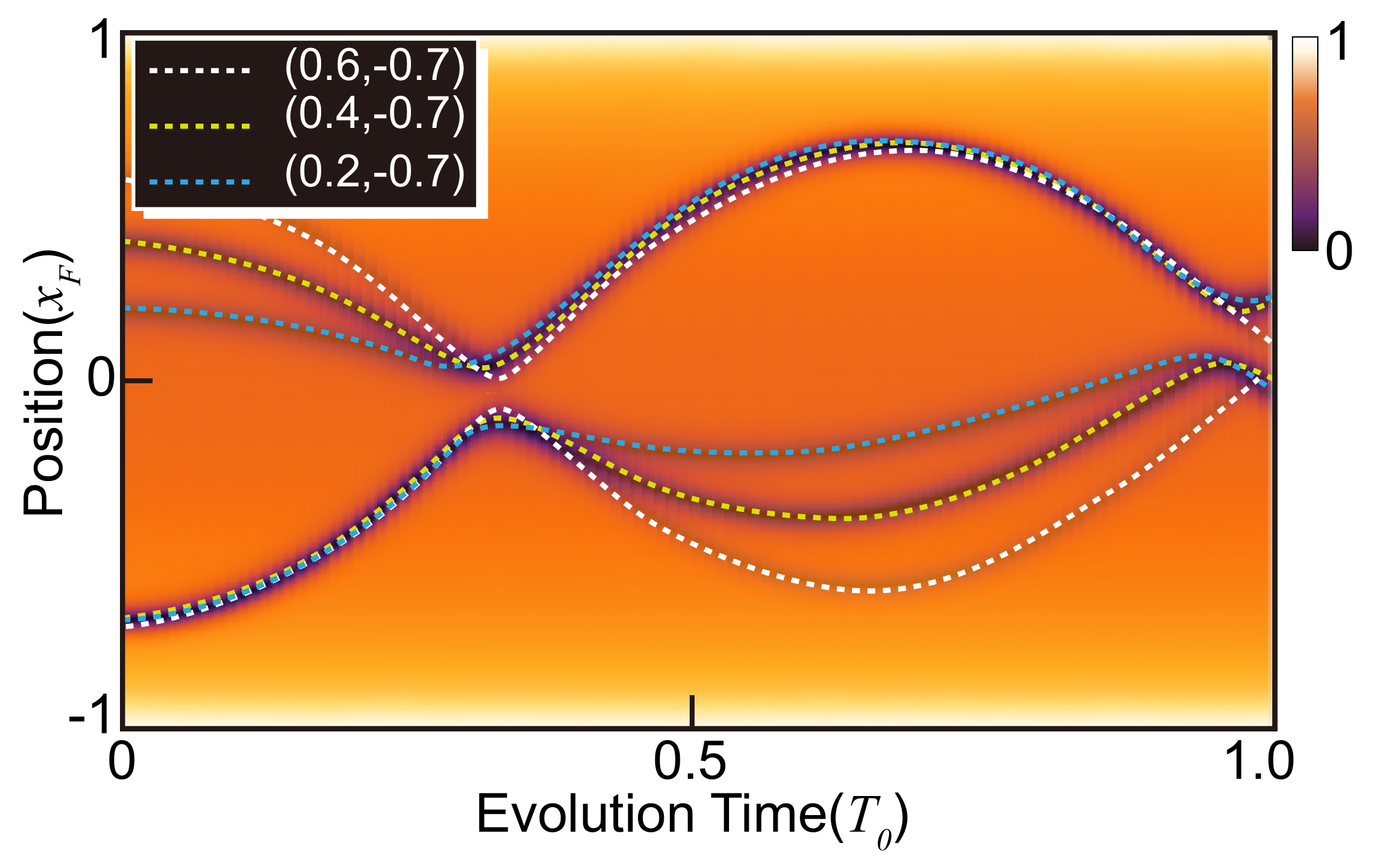}\\
\caption{(color online) Space-time plot of the asymmetrical collisions with two solitons prepared at ($-0.7x_{F},0.2x_{F}$), ($-0.7x_{F},0.4x_{F}$) and ($-0.7x_{F},0.6x_{F}$), respectively. The quasiclassical predictions of the soliton’s trajectories are highligthed with dashed lines, which show a good agreement with the numerical results. }
\label{Fig: trajectories}
\end{figure}

\textit{Quasiclassical Analysis of Soliton Collision.}--
%====[Semiclassical Analysis-01]====
Dark soliton appear as wave packet that preserve its amplitude and shape during its propagation and even persist unchanged with MZMs in the colliding process, therefore being attributed as a particle-like character.
%====[Semiclassical Analysis-01]====
In our study, the width of the Fermi cloud in the axial direction is much larger compared to the size of the soliton.
 %====[Semiclassical Analysis-01]====
 Thus, under the local density approximation, the soliton can be treated as a macroscopic particle at coordinates $q_{i}$ with $p_{i}$ being generalized momentum.
%====[Semiclassical Analysis-01]==== 
In our study, the overlapping between MZMs adjusts to the soliton's motion and plays an important role in the soliton collision.
%====[Semiclassical Analysis-01]==== 
According to the energy splitting from zero $\varpropto \pm \exp(-|q_{1}-q_{2}|/\xi_{M})$ induced by the Majorana wave-function overlapping\cite{Cheng-2009}, their mutual repulsive potential can be assumed reasonably as $V(q_{1},q_{2})=Ae^{-|q_{1}-q_{2}|/\sqrt{\xi_{M_{1}}\xi_{M_{2}}}}$, where $A$ is the interaction intensity to be determined.
%====[Semiclassical Analysis-01]====
The soliton matter interaction is negligible and we may finally obtain the Hamiltonian of the double solitons by evaluating the kinetic, trap and repulsive interaction energy terms in the absence of dissipation,
\begin{equation}
H(q_{i},p_{i})=\sum_{i=1,2}\frac{p_{i}^2}{2m_{\mathrm{I}}^{i}}+\sum_{i=1,2}\frac{1}{2}m_{\mathrm{S}}^{i}\omega^2q_{i}^2+V(q_{1},q_{2}).
\label{Semiclassical Hamiltonian}
\end{equation}
where $m_{\mathrm{I}}$ is the inertial mass and $m_{\mathrm{S}}=N_{\mathrm{S}}m$ bare soliton mass, which can be determined in the snaking process of single soliton (as elaborated on in Supplementary Material).
%====[Semiclassical Analysis-01]====
With the initial state $\left\{q_{i}(t=0)=x_{i},p_{i}(t=0)=0\right\}$, following the Hamiltonian equation of motion $\dot{q}_{i}=\partial H/\partial p_{i}$, $\dot{p}_{i}=-\partial H/\partial q_{i}$, the interaction intensity is fixed to be $A=0.0137E_{F}$ compared with the soliton trajectory based on the full numerical BdG simulation in Fig.(\ref{Fig: Asymmetrical Collision}).
%====[Semiclassical Analysis-01]====
The numerical results of the trajectories of the soliton centers and the time dependences of the soliton's coordinates originating from Eq.\ref{Semiclassical Hamiltonian}, are presented in Fig.(\ref{Fig: trajectories}) with different initial states.
%====[Semiclassical Analysis-01]====
All the analytical results are in good agreement with the numerical data.
%The solid lines in Fig. (\ref{Fig:Soliton Collision}a) presents the analytical solution described in Eq. \ref{Semiclassical Hamiltonian}, to agree well with the numerical data.
%====[Semiclassical Analysis-01]====

\begin{figure}[t]
\centering
\includegraphics[width=0.49\textwidth]{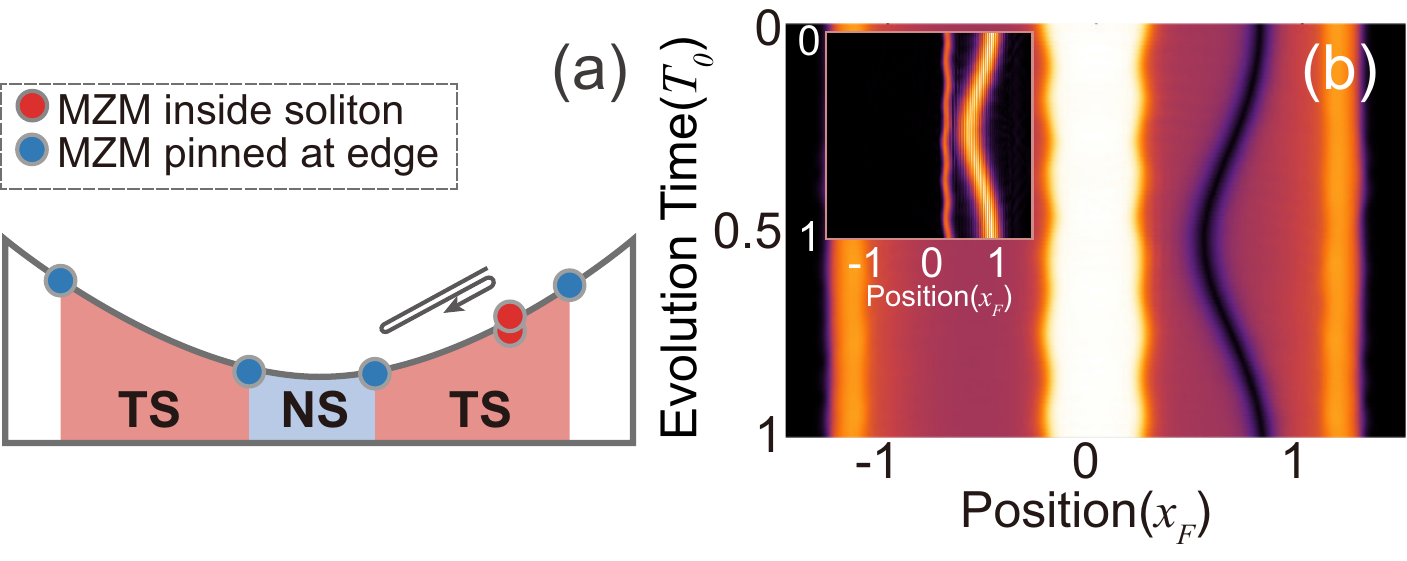}\\
\caption{(color online.) (a) A rough configuration of the system in a harmonic trap within the local-density approximation. From the center of the trap to the wings the phases are normal superfluid (NS) phase and topological superfluid (TS) respectively.
(b) shows the dynamical evolutions of soliton in the TS region.
}
\label{Fig: PTS}
\end{figure}

\textit{Collision with phase-edge Majorana states.}--
%====[Pinned-01]====
%and for concreteness
Then, in order to check the above-made statements and extend our works into the collision between MZMs within different topological defects, we perform numerical simulations of the collision between soliton-induced MZMs and one pinned at the phase edge.
%Then, in order to confirm the repulsion between MZMs, we perform numerical simulations of collision between soliton-induced MZM and one pinned at the phase edge.
%====[Pinned-01]====
Due to the harmonic potential geometry, in a suitable parameter regime, a mixed phase emerges, consisting of a standard normal superfluid at the center and a topological superfluid at the two edges of the trap\cite{Liu-2012}, which is shown in Fig.(\ref{Fig: PTS}a).
%====[Pinned-01]====
A dark soliton was set in the topological region and accelerated by the trapping potential towards to the phase edge, where a MZM is prepared immovablely.
%====[Pinned-01]====
%There are six Majorana states, two hosted in the soliton core and four pinned at the phase edges.
%====[Pinned-01]====
As shown in Fig.(\ref{Fig: PTS}b), the soliton slow down as it approaches the phase edge at $0.2x_{F}$, and then turn back, performing a complete oscillation (a virtual phase edge reflection) inside the topological region.
%====[Pinned-01]====
Or in other words, Majorana states hosted in the soliton has been limited in the span between two fixed MZMs, reminiscent of what has been observed in the soliton collision.
%====[Pinned-01]====
The soliton behavior is generated by the conjunction of the trapping effect and repulsive interaction between Majorana states, which is an reconfirmation of the repulsive interaction of MZMs.
%====[Pinned-02]====
%Based on the above study, the elasticity of the collision between Majorana states in different defects (such as dark soliton, phase edge) can therefore be explained only via the concept of the repulsive effect between Majorana states.
%Furthermore, we have explored how these repulsive effects can be exploited to actively manipulate and control the propagation of Majorana states.

\textcolor{magenta}
{
%====[Pinned-02]====
%Having demonstrated that there is an effective repulsion between Majorana states, we now explore how these repulsive effects can be exploited to actively manipulate and control the propagation of Majorana states.
%====[Pinned-02]====
%As shown in Fig.(\ref{Fig: Zeeman field}a), the position of the phase edge depends critically on the Zeeman field.
%====[Pinned-02]====
%Therefore, by adiabatically increasing or decreasing the Zeeman field, the inner two MZMs on the phase edge will move towards or away from the trap center, even disappear.
%====[Pinned-02]====
%The position of the outer two MZMs, however, is less affected by the Zeeman field.
%====[Pinned-02]====
%As shown in Fig.(\ref{Fig: Zeeman field}b), 
}

%\begin{figure}[t]
%\centering
%\includegraphics[width=0.49\textwidth]{fig-007.eps}\\
%\caption{(color online.) (a) A rough configuration of the system in a harmonic trap within the local-density approximation. From the %center of the trap to the wings the phases are normal superfluid (NS) phase and topological superfluid (TS) respectively.
%(b) shows the dynamical evolutions of soliton in the TS region.
%(c) The position of the phase edge depends on the Zeeman field.
%}
%\label{Fig: Zeeman field}
%\end{figure}

%\textit{Boosting and Manipulating Soliton with Majorana states.}--
%====[Pinned-02]====
%Having demonstrated that there is an effective repulsive interaction between Majorana states, we now explore how these repulsive effects can be exploited to actively manipulate and control the propagation of Majorana states.
%====[Pinned-02]====
%As shown in Fig. (\ref{Fig: Boosting and Manipulating}a), the position of the phase edge depends critically on the Zeeman field.
%====[Pinned-02]====
%Therefore, by adiabatically increasing or decreasing the Zeeman field, the inner two MZMs on the phase edge will move towards or away from the trap center, even disappear.
%====[Pinned-02]====
%The position of the outer two MZMs, however, is less affected by the Zeeman field.
%====[Pinned-02]====
%As shown in Fig. (\ref{Fig: Boosting and Manipulating}b), (we should add some statements about the changes in the propagation of the Majorana soliton and the moving magnetic impurity to boost the resting soliton).

\textit{Conclusion.}--
%====[Conclusion]====
To summarize, our study shows that the zero energy splitting, induced by the overlapping of inter-soliton Majorana wave-functions upon soliton collision, generates an effective repulsive force for Majorana states, which in turn protected themselves against into bulk excitation.
%====[Conclusion]====
A complete elastic collision takes place between solitons with Majorana states, which do not penetrate each other but instead repel without any loss of energy.
%====[Conclusion]====
Remarkably, our quasi-particle analysis can fully explain the numerical findings and provide a complete description of these anomalous behaviors with Majorana states.
%====[Conclusion]====
Additionally, we perform the investigation of the collisional mechanism between Majorana states within different topological defects to confirm our conjecture.
%====[Conclusion]====
Our research provides new insights into the features of Majorana fermions, and we envision that the robustness in the collisions of Majorana states could be utilized in the topological quantum computing with a network of Majorana qubits.

%To summarize, our study shown that effective repulsive interactions can occur between two Majorana solitons.
%====[Conclusion]====
%The Majorana modes split due to the inter-soliton approaching, and always stays in the gap, which in turn creates a protecting barrier for solitons that block their propagation.
%====[Conclusion]====
%A complete elastic collision is taken place with the energy conservation, which is closely related to the Majorana topological protected nature. 
%====[Conclusion]====
%Furthermore, we perform the investigation of the collisional mechanism between MZMs in soliton and ones fixed at the phase edge and magnetic impurity, to confirm our conjecture.
%====[Conclusion]====
%Remarkably, our quasi-particle analysis can fully explain the numerical findings and provide a complete description of these anomalous behaviors with Majorana states.
%====[Conclusion]====
%We envision that the repulsive effect between Majorana states could be used for control and manipulation of the Majorana qubits for fault-tolerant topological quantum computations.
%These unusual properties of dark solitons with Majorana quasiparticles may open an alternative way to detect and manipulate Majorana fermions for fault-tolerant topological quantum computations.

We thank  An-chun Ji, Qing Sun and Changan Li for useful discussions. This work was supported by the foundation of Zhejiang Province Natural Science under Grant No. LQ20A040002 and JL acknowledges support from National Natural Science Foundation of China under Project 11774317. The numerical calculations in this paper have been done on the super-computing system in the Information Technology Center of Westlake University.

 \end{document}

\title{Supplemental Material for ``Collisions of Majorana Zero Modes"}
\author{Liang-Liang Wang}
%\email{wangliangliang@westlake.edu.cn}
\affiliation{School of Science, Westlake University, 18 Shilongshan Road, Hangzhou 310024, Zhejiang Province, China}
\affiliation{Institute of Natural Sciences, Westlake Institute for Advanced Study, 18 Shilongshan Road, Hangzhou 310024, Zhejiang Province, China}
\author{Wenjun Shao}
%\email{wangliangliang@westlake.edu.cn}
\affiliation{School of Science, Westlake University, 18 Shilongshan Road, Hangzhou 310024, Zhejiang Province, China}
\affiliation{Institute of Natural Sciences, Westlake Institute for Advanced Study, 18 Shilongshan Road, Hangzhou 310024, Zhejiang Province, China}
\author{Jian Li}
\email{lijian@westlake.edu.cn}
\affiliation{School of Science, Westlake University, 18 Shilongshan Road, Hangzhou 310024, Zhejiang Province, China}
\affiliation{Institute of Natural Sciences, Westlake Institute for Advanced Study, 18 Shilongshan Road, Hangzhou 310024, Zhejiang Province, China}

\date{{\small \today}}
\maketitle

In this Supplementary Material, we present the self-consistent time-splitting technique for the TDBdG equation, and show the asymmetric collision between solitons.

\subsection{The self-consistent time-splitting technique for the TDBdG equation}
\begin{figure}[t]
\centering
\includegraphics[width=0.98\textwidth]{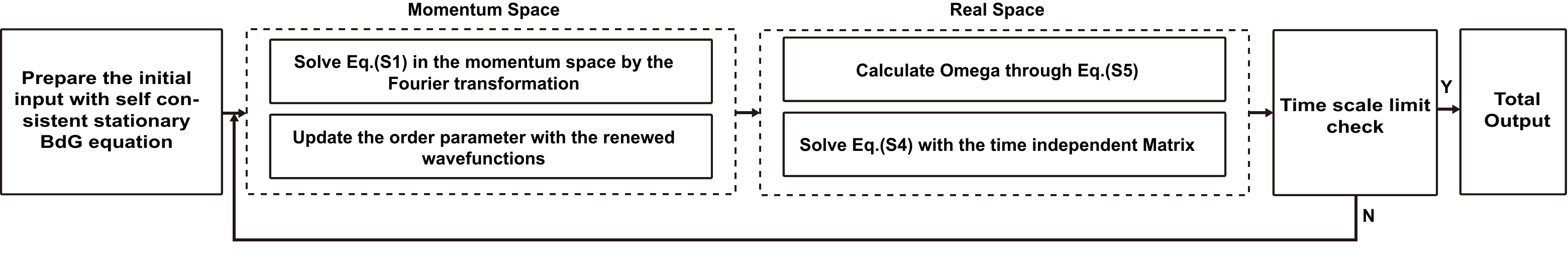}\\
\caption{(color online) 
Schematic flow chart for the self-consistent time-splitting technique of the TDBdG equation.
}
\label{Fig: Numerical Process}
\end{figure}

In Fig. (\ref{Fig: Numerical Process}), we sketch the self-consistent time-splitting technique for the TDBdG equation.
%%=== Part-01 ===
We prepared the initial states by solving the self-consistent stationary BdG equation numerically.
%%=== Part-01 ===
The following iterative algorithm is taken: one starts with an initial guess with the phase kink for the gap parameter, and then calculates the $u_{\sigma,\eta}$'s and $v_{\sigma,\eta}$'s by the diagonalization of the Hermitian BdG matrix.
%%=== Part-01 ===
The corresponding $\Delta(x)$ is calculated by using the self-consistent order parameter equation, and iterated until pre-set convergence.
%%=== Part-01 ===
We have to truncate the summation over energy levels $\eta$. This is done by introducing a high energy cut-off $E_{c}$.
%%=== Part-01 ===
Once the stationary state is calculated, one moves to the solution of the time dependent BdG equation to simulate the soliton dynamics.
%%=== Part-01 ===
We consider the standard splitting technique with self-consistency that approximates the evolution during a small time step $dt$.
%%=== Part-01 ===
By firstly evolving the $[u_{\uparrow, \eta}(x,t_{i})$,$u_{\downarrow, \eta}(x,t_{i})$, $v_{\uparrow, \eta}(x,t_{i})$, $v_{\downarrow, \eta}(x,t_{i})]$ $\Rightarrow$ $[u^{'}_{\uparrow, \eta}(x,t_{i})$,$u^{'}_{\downarrow, \eta}(x,t_{i})$, $v^{'}_{\uparrow, \eta}(x,t_{i})$, $v^{'}_{\downarrow, \eta}(x,t_{i})]$ with the kinetic energy and spin-orbit coupling during $dt$,
\begin{equation}
i\hbar\frac{\partial}{\partial t}
\begin{bmatrix}
u_{\uparrow,\eta}\\
u_{\downarrow, \eta} \\
v_{\uparrow, \eta}\\
v_{\downarrow, \eta}\\
\end{bmatrix}
=
\begin{bmatrix}
\hat{k}_{x}^2/2 & -i\lambda \hat{k}_{x} & 0 & 0\\
i\lambda \hat{k}_{x} & \hat{k}_{x}^2/2 & 0 & 0\\
0 & 0 & - \hat{k}_{x}^2/2 & i\lambda \hat{k}_{x} \\
0 & 0 & -i\lambda \hat{k}_{x} & - \hat{k}_{x}^2/2\\
\end{bmatrix}
\begin{bmatrix}
u_{\uparrow,\eta}\\
u_{\downarrow, \eta} \\
v_{\uparrow, \eta}\\
v_{\downarrow, \eta}\\
\end{bmatrix}
,
\end{equation}
the above equation can be calculated in the momentum space by the Fourier transform, where
\begin{equation}
\begin{split}
\begin{bmatrix}
u_{\uparrow,\eta}^{'}\\
u_{\downarrow,\eta}^{'}\\
\end{bmatrix}
&=
\frac{1}{2}
\begin{bmatrix}
-i & i  \\
1 & 1 \\
\end{bmatrix}
\mathbf{ifft}\{
\begin{bmatrix}
A_{1} & 0 \\
0 & A_{2}\\
\end{bmatrix}
\begin{bmatrix}
 i & 1\\
 -i & 1 \\
\end{bmatrix}
\mathbf{fft}
\{
\begin{bmatrix}
u_{\uparrow,\eta}\\
u_{\downarrow,\eta}\\
\end{bmatrix}
\}
\}
\\
\begin{bmatrix}
v_{\uparrow,\eta}^{'}\\
v_{\downarrow,\eta}^{'}\\
\end{bmatrix}
&=
\frac{1}{2}
\begin{bmatrix}
i & -i  \\
1 & 1 \\
\end{bmatrix}
\mathbf{ifft}\{
\begin{bmatrix}
B_{1} & 0 \\
0 & B_{2}\\
\end{bmatrix}
\begin{bmatrix}
 -i & 1\\
 i & 1 \\
\end{bmatrix}
\mathbf{fft}
\{
\begin{bmatrix}
v_{\uparrow,\eta}\\
v_{\downarrow,\eta}\\
\end{bmatrix}
\}
\}
\\
\end{split}
,
\end{equation}
with the matrix elements are $A_{1}=e^{-idt(\hat{k}^2/2+\lambda \hat{k})}$, $A_{2} =e^{-idt(\hat{k}^2/2-\lambda \hat{k})}$, $B_{1} =e^{-idt(-\hat{k}^2/2+\lambda \hat{k})}$, $B_{2} =e^{-idt(-\hat{k}^2/2-\lambda \hat{k})}$.
%%=== Part-01 ===
After the renewed wavefunctions are obtained, we then update the order parameter through
\begin{equation}
\Delta^{'}(x, t_{i}) = -g_{\mathrm{1D}}\sum_{\eta}[u^{'}_{\uparrow, \eta}(x,t_{i})v_{\downarrow, \eta}^{\ast'}(x,t_{i})f(-E_{\eta})+u^{'}_{\downarrow, \eta}(x,t_{i})v_{\uparrow, \eta}^{\ast'}(x,t_{i})f(E_{\eta})]
\end{equation}
for the next procedure.
%%=== Part-01 ===
The $[u^{'}_{\uparrow, \eta}(x,t_{i})$, $u^{'}_{\downarrow, \eta}(x,t_{i})$, $v^{'}_{\uparrow, \eta}(x,t_{i})$ , $v^{'}_{\downarrow, \eta}(x,t_{i})]$ $\Rightarrow$ $[u_{\uparrow, \eta}(x,t_{i}+dt)$, $u_{\downarrow, \eta}(x,t_{i}+dt)$, $v_{\uparrow, \eta}(x,t_{i}+dt)$ , $v_{\downarrow, \eta}(x,t_{i}+dt)]$ are evolved with the spatial part of the matrix during $dt$, where
\begin{equation}
i\hbar\frac{\partial}{\partial t}
\begin{bmatrix}
u^{'}_{\uparrow,\eta}\\
u^{'}_{\downarrow,\eta}\\
v^{'}_{\uparrow,\eta}\\
v^{'}_{\downarrow,\eta}\\
\end{bmatrix}
=
\begin{bmatrix}
C_{\uparrow} & 0 & 0 & \Delta(x,t)\\
0 & C_{\downarrow} & -\Delta(x,t) & 0\\
0 & -\Delta^{\ast}(x,t) & -C_{\uparrow} & 0 \\
\Delta^{\ast}(x,t) & 0 & 0 &  -C_{\downarrow}\\
\end{bmatrix}
\begin{bmatrix}
u^{'}_{\uparrow,\eta}\\
u^{'}_{\downarrow,\eta}\\
v^{'}_{\uparrow,\eta}\\
v^{'}_{\downarrow,\eta}\\
\end{bmatrix}
,
\label{real-space}
\end{equation}
with $C_{\uparrow} = V(x) -\mu -h_{z}$, $C_{\downarrow} = V(x) -\mu +h_{z}$ and $ \Delta(x,t_{i})=\Delta^{'}(x,t_{i})$. 
%%=== Part-01 ===
As we all know, $\Delta(x,t)$ is not a fixed but time dependent order parameter. 
%%=== Part-01 ===
According to the self-consistent condition, it exactly preserves the unitary of the wavefunctions during the entire evolution, but taking a fixed value of $\Delta(x,t_{i})$ during the second step of the evolution obviously breaks the indispensable self-consistency between $\Delta$ and $u_{\sigma,\eta}$, $v_{\sigma,\eta}$. 
%%=== Part-01 ===
Therefore the total number of particles is conserved to first order in the time step but not to all orders in $dt$. 
%%=== Part-01 ===
We go beyond the standard splitting method by resorting the self-consistency in the evolution during $dt$.
%%=== Part-01 ===
With a simple analysis
\begin{equation}
\begin{split}
i\hbar\frac{\partial}{\partial t}\Delta(x,t) & = -i\hbar\frac{\partial}{\partial t}\sum_{\eta}g_{\mathrm{1D}}\left[u_{\uparrow,\eta}(x,t)v_{\downarrow, \eta}^{\ast}(x,t)f(-E_{\eta})+u_{\downarrow, \eta}(x,t)v_{\uparrow, \eta}^{\ast}(x,t)f(E_{\eta})\right]\\
& =\left[2(V(x)-\mu)-g_{\mathrm{1D}}\sum_{\eta}\left(|v_{\downarrow, \eta}|^{2}-|u_{\uparrow, \eta}|^{2}\right)f(-E_{\eta})+\left(|v_{\uparrow, \eta}|^{2}-|u_{\downarrow, \eta}|^{2}\right)f(E_{\eta})\right]\Delta(x,t),
\end{split}
\end{equation}
we fortunately find that $\Omega\equiv 2(V(x)-\mu)-g_{\mathrm{1D}}\sum_{\eta}(|v_{\eta}|^{2}-|u_{\eta}|^{2})f(-E_{\eta})+\left(|v_{\uparrow, \eta}|^{2}-|u_{\downarrow, \eta}|^{2}\right)f(E_{\eta})$ is time independent with zero temperature. 
%%=== Part-01 ===
Thus the Eq. \ref{real-space} can be transformed with $u_{\sigma,\eta}(x,t)=U_{\sigma,\eta}(x,t)e^{-i\Omega(x)t/2}$ and $v_{\sigma,\eta}(x,t)=V_{\sigma,\eta}(x,t)e^{i\Omega(x)t/2}$ and then  evolve the $\left[U_{\sigma,\eta}(x,t),V_{\sigma,\eta}(x,t)\right]$ with the time independent $4\times 4$ matrix.
%%=== Part-01 ===
We sketch the self-consistent time-splitting procedure in Fig. (\ref{Fig: Numerical Process}).
%%=== Part-01 ===
It is clear that, about 1000 $\left\{u_{\uparrow},u_{\downarrow},v_{\uparrow},v_{\downarrow}\right\}$' have to be simultaneously solved in the process, the computing task is rather demanding.

\begin{figure}[b]
\centering
\includegraphics[width=0.85\textwidth]{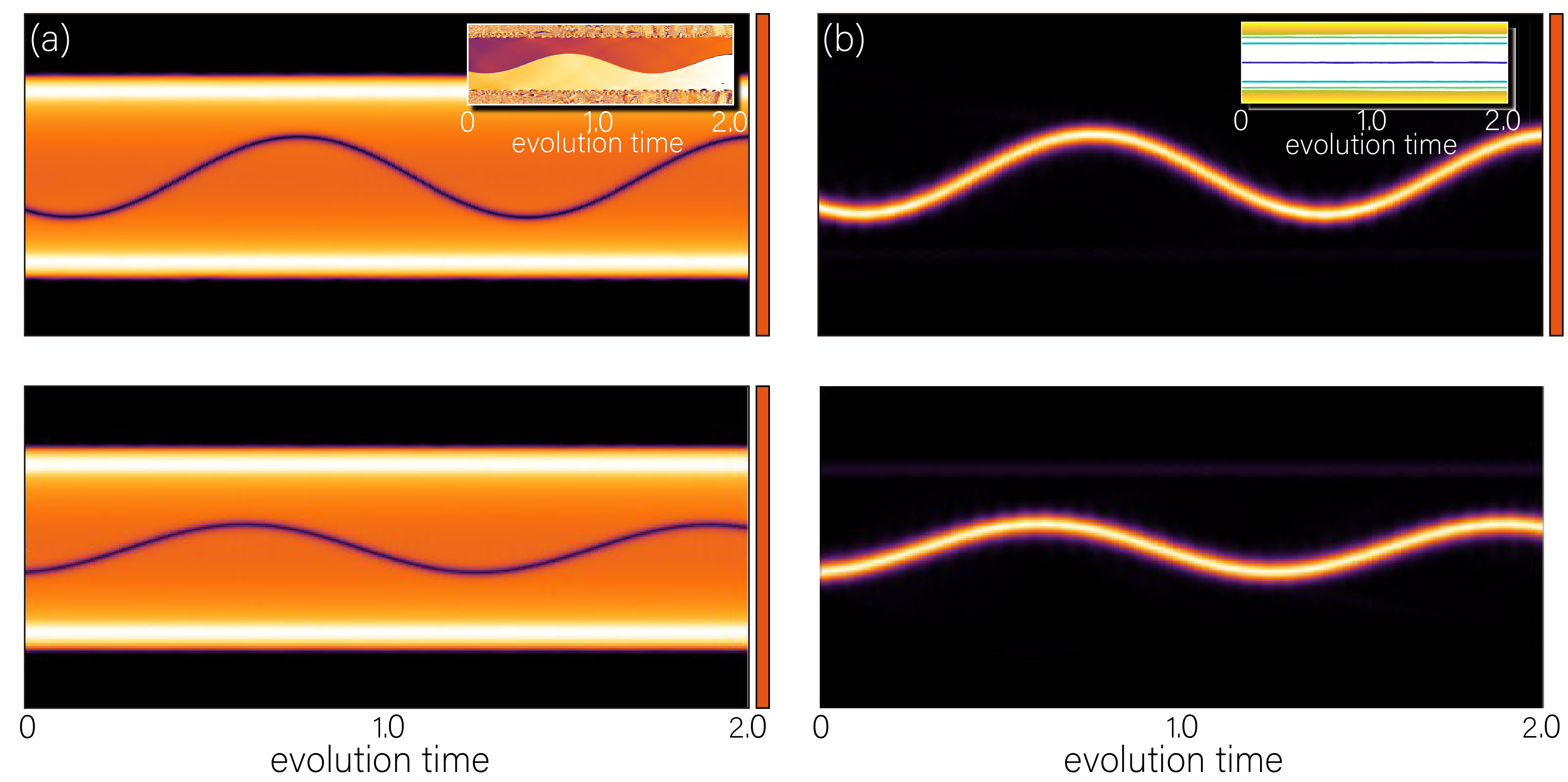}\\
\caption{(color online) 
Schematic flow chart for the self-consistent time-splitting technique of the TDBdG equation.
}
\label{Fig: Single Soliton Oscillation}
\end{figure}

\subsection{Quasi-classical properties of the soliton in the system}
%%=== Part-02 ===
Here we consider the single soliton in a trapped Fermi superfluid, which preform a stable oscillation as a classical particle.
%%=== Part-02 ===
Thus, under the local density approximation (LDA), the soliton can be treated as a macroscopic particle at position $X$ with $V = dX/dt$ being the soliton velocity. 
%%=== Part-02 ===
For the dark soliton with phase-jump, it can be seen as a surface (or domain wall) in the Fermi gas. 
%%=== Part-02 ===
The associated surface tension (or the soliton energy) $E_{s}(\mu, V^2)\equiv E_{\mathrm{with soliton}}(\mu, V^2)-E_{\mathrm{without soliton}}(\mu)$ can be a function of the chemical potential $\mu$ and $V^2$. 
%%=== Part-02 ===
Under the LDA, in the harmonic trap, we can say that
%%=== Part-02 ===
$\mu(X)=\mu-U(X)$.
%%=== Part-02 ===
According to the numerical results, the dark soliton with Majorana zero modes preserve their amplitude and shape during propagation and even persist elastic from collisions with one another. 
%%=== Part-02 ===
The energy conversion in the absence of dissipation gives
\begin{equation}
\frac{\partial E_{s}}{\partial t}=\frac{\partial E_{s}}{\partial \mu(X)}\cdot \frac{d \mu(X) }{dX}\cdot \frac{dX}{dt}+\frac{\partial E_{s}}{\partial V^2}\cdot \frac{dV^2}{dV}\cdot \frac{dV}{dt}=0.
\end{equation}
We define the inertial mass and the soliton particle number
\begin{equation}
 m_{I}=2\frac{\partial E_{s}}{\partial V^2}, N_{s}=-\frac{\partial E_{s}}{\partial \mu(X)}=\int(-\frac{\partial H_{s}}{\partial \mu(X)})-(-\frac{\partial H_{0}}{\partial \mu(X)})dX=\int [n_{s}(x)-n_{0}(x)]dx
,
\end{equation}
%%=== Part-02 ===
in which $n_{s}(x)$ is the density with dark soliton and $n_{0}(x)$ is the density without the dark soliton. 
%%=== Part-02 ===
The quantity $N_{s}$ is the deficit of the particles associated with the depression in the density at the soliton position.
%%=== Part-02 ===
Note that, for dark solitons typically, $m_{I}<0$ and $N_{s}<0$.
%%=== Part-02 ===
Hence, the Equation of the energy conservation gives
\begin{equation}
m_{I}\frac{dV}{dt}=-N_{s}\frac{dU}{dX}=-N_{s}m\omega_{x}^2X.
\end{equation}
%%=== Part-02 ===
For small amplitude oscillations, the soliton inertial mass is
$m_{I}=N_{s}m[{T_{s}}/{T_{x}}]^2$,
where $T_{x}=2\pi/\omega_{x}$ and $T_{s}$ is the soliton period, which can be obtained by the numerical data.
%%=== Part-02 ===
Via the result shown in Fig. (\ref{Fig: Single Soliton Oscillation}), we can determine the value of $T_{s}$ and $N_{s}$, then we can give the soliton inertial mass by the above relationship.